\documentclass[letter]{emulateapj}
\usepackage{graphicx}
\usepackage{natbib}
\usepackage{float}
\usepackage{color}

\newcommand{\gcc}{\ensuremath{\mathrm{g}\,\mathrm{cm}^{-3}}}

\newcommand{\nuc}[2]{\ensuremath{\mathrm{^{#1}#2}}}
\newcommand{\ions}[2]{#1\,{\sc #2}}

\newcommand{\msun}{\ensuremath{\mathrm{M}_\odot}}

\def\mch{$M_\mathrm{Ch}$}
\def\dm15{$\Delta m_{15}(B)$}
\def\lesssim{\mathrel{\hbox{\rlap{\hbox{\lower4pt\hbox{$\sim$}}}\hbox{$<$}}}}

\def\gtrsim{\mathrel{\hbox{\rlap{\hbox{\lower4pt\hbox{$\sim$}}}\hbox{$>$}}}}

\def\aj{AJ}%
\def\araa{ARA\&A}%
\def\apj{ApJ}%
\def\apjl{ApJL}%
\def\aap{A\&A}%
\def\mnras{MNRAS}%
\def\nat{Nature}%
\shortauthors{Kromer et al.}

\begin{document}
\title{SN~2010lp---a Type Ia supernova from a violent merger of two carbon--oxygen White Dwarfs}

\author
{
  M.~Kromer\altaffilmark{1,2}, R.~Pakmor\altaffilmark{3}, 
  S.~Taubenberger\altaffilmark{1}, G.~Pignata\altaffilmark{4},
  M.~Fink\altaffilmark{5}, F.~K.~R\"opke\altaffilmark{5},
  I.~R.~Seitenzahl\altaffilmark{5,1},
  S.~A.~Sim\altaffilmark{6} \& W.~Hillebrandt\altaffilmark{1}
}

\altaffiltext{1}
{Max-Planck-Institut f\"ur Astrophysik,
  Karl-Schwarzschild-Str.\ 1, 
  85748 Garching, Germany} 
\altaffiltext{2}
{The Oskar Klein Centre, 
  Stockholm University, AlbaNova, 
  SE-106 91 Stockholm, Sweden}
\altaffiltext{3}
{Heidelberger Institut f\"{u}r Theoretische Studien, 
  Schloss-Wolfs\-brunnen\-weg 35, 
  69118 Heidelberg, Germany} 
\altaffiltext{4}
{Departamento de Ciencias Fisicas, Universidad Andres Bello,
  Avda.\ Republica 252, 
  Santiago, Chile} 
\altaffiltext{5}
{Institut f\"ur Theoretische Physik und Astrophysik,
  Universit\"at W\"urzburg,
  Emil-Fischer-Str.\ 31,
  97074 W\"urzburg, Germany} 
\altaffiltext{6}
{Astrophysics Research Centre, 
 School of Mathematics and Physics, 
 Queen's University Belfast, 
 Belfast BT7 1NN, UK}

\date{Received ; accepted}

\begin{abstract}
  SN~2010lp is a subluminous Type Ia supernova (SN~Ia) with 
  slowly-evolving lightcurves. Moreover, it is the only 
  subluminous SN~Ia observed so far 
  that shows narrow emission lines of [\ions{O}{i}] in late-time 
  spectra, indicating unburned oxygen close to the centre of the ejecta.
  Most explosion models for SNe~Ia cannot explain the narrow 
  [\ions{O}{i}] emission. Here, we present hydrodynamic explosion
  and radiative transfer calculations showing that the violent 
  merger of two carbon--oxygen white dwarfs of 0.9 and 0.76 \msun, 
  respectively, adequately reproduces the early-time observables of
  SN~2010lp. Moreover, our model predicts oxygen close to the centre
  of the explosion ejecta, a pre-requisite for narrow [\ions{O}{i}] 
  emission in nebular spectra as observed in SN~2010lp.
\end{abstract}


\maketitle\section{Introduction}
\label{sec:introduction}

Type Ia supernovae (SNe~Ia) form a relatively homogeneous class of
objects. It is widely accepted that they result from thermonuclear 
explosions of carbon--oxygen (CO) white dwarfs (WDs) in binary systems
(see e.g.\ \citealt{hillebrandt2000a}). However, in the absence
of direct progenitor detections \citep{li2011b, bloom2012a} both 
the exact nature of the progenitor systems (double- versus
single-degenerate) and the details of the explosion mechanism are not 
well known \citep{hillebrandt2000a}. 
There is now indirect 
observational evidence \citep{patat2007a, li2011b, bloom2012a, 
dilday2012a, hernandez2012a, schaefer2012a, shappee2013a} 
that more than one progenitor type may be responsible for the bulk
of normal SNe~Ia \citep[e.g.][for a review]{hillebrandt2013a}.

A promising method to investigate the nature of SNe~Ia is the
comparison of theoretical explosion models to observed SN lightcurves 
and spectra. This has been used extensively in the past
(e.g.\ \citealt{hoeflich1996a,kasen2009a,sim2010a}).
However, it can be very difficult to discriminate between different 
progenitor models for normal
SNe~Ia from optical lightcurves and spectra alone \citep{roepke2012a}. 
Nevertheless, specific progenitor and explosion models have been
identified for some of the peculiar sub-classes of SNe~Ia, like 
e.g.\ the subluminous 2002cx-like SNe \citep{jordan2012b,kromer2013a}.

In a companion paper, Pignata et al.\ (in prep.) present observations
of SN~2010lp, a peculiar subluminous SN~Ia in NGC 1137 (distance
modulus $\mu=33.03$\,mag, redshift $z=0.010$, $E(B-V)=0.21$\,mag
assumed). At early times this SN 
has similar brightness and spectra to the subluminous 1991bg-like 
SNe \citep{filippenko1992b, leibundgut1993a} and, like those, does not 
show double-peaked $I$-band lightcurves. However, the lightcurve 
evolution of SN~2010lp is much slower than in 1991bg-like SNe. 
Pignata et al.\ (in prep.) e.g.\ derive a 
\dm15 of 1.24 for SN~2010lp, while typical 1991bg-like SNe have
a \dm15 of ${\sim}1.9$ \citep{taubenberger2008a}. 

Moreover, narrow [\ions{O}{i}] emission has been identified in a 
nebular spectrum of SN~2010lp \citep{taubenberger2013b}, indicating 
a concentration of unburned 
material near the centre of the ejecta. This is very 
challenging for most current explosion scenarios of SNe~Ia discussed
in the literature. Neither Chandrasekhar-mass (\mch) delayed detonations
\citep[e.g.][]{seitenzahl2013a} nor sub-\mch\ double detonations 
and edge-lit detonations \citep{nomoto1982b,livne1990a,fink2010a}
are expected to leave O in the central ejecta, since the 
densities there are so high that almost all the fuel is completely
burned to iron-group elements (IGEs). In contrast, in 
\mch-deflagrations the turbulent burning leads to a well-mixed 
ejecta structure, i.e.\ O is present down to the centre of the ejecta
(e.g.\ \citealt{ma2013a,fink2013a}). As shown by
\citet{kozma2005a}, [\ions{O}{i}] emission can be present in 
such models. However, since O and IGEs are abundant 
over a wide velocity range, the predicted [\ions{O}{i}] feature is 
broad, unlike the narrow feature observed in SN~2010lp.
\citet{taubenberger2013b} pointed out that an ejecta structure
similar to that of the violent merger of a 1.1 and a 0.9 \msun
WD as presented in \citet{pakmor2012a} might produce narrow 
[\ions{O}{i}] emission 
in a small region close to the centre of the ejecta. The 
particular 1.1+0.9~\msun\ merger of \citet{pakmor2012a}, however,
is significantly too bright to be a good match to SN~2010lp. 

Here, we present a hydrodynamic simulation of the violent merger of 
two CO WDs of 0.9 and 0.76\,\msun, respectively. By calculating 
synthetic observables from radiative transfer simulations, we find
that this model reproduces the observed early-time properties of 
SN~2010lp extremely well. Moreover, the explosion ejecta contain
O near the centre, which might potentially lead to [\ions{O}{i}] 
emission in nebular spectra as observed in SN~2010lp.

The Letter is organized as follows. In Section~\ref{sec:merger} we 
describe the progenitor system and the hydrodynamic explosion 
simulation of our model. In Section~\ref{sec:nucleosynthesis} we 
present the detailed chemical composition of the explosion ejecta, 
which we then use in Section~\ref{sec:observables} to obtain 
synthetic observables and compare them with 
SN~2010lp. In Section~\ref{sec:discussion} we discuss our findings 
before drawing conclusions in Section~\ref{sec:conclusions}.

\section{Merger and Explosion}
\label{sec:merger}

\begin{figure*}
   \centering
   \includegraphics[width=0.95\linewidth]{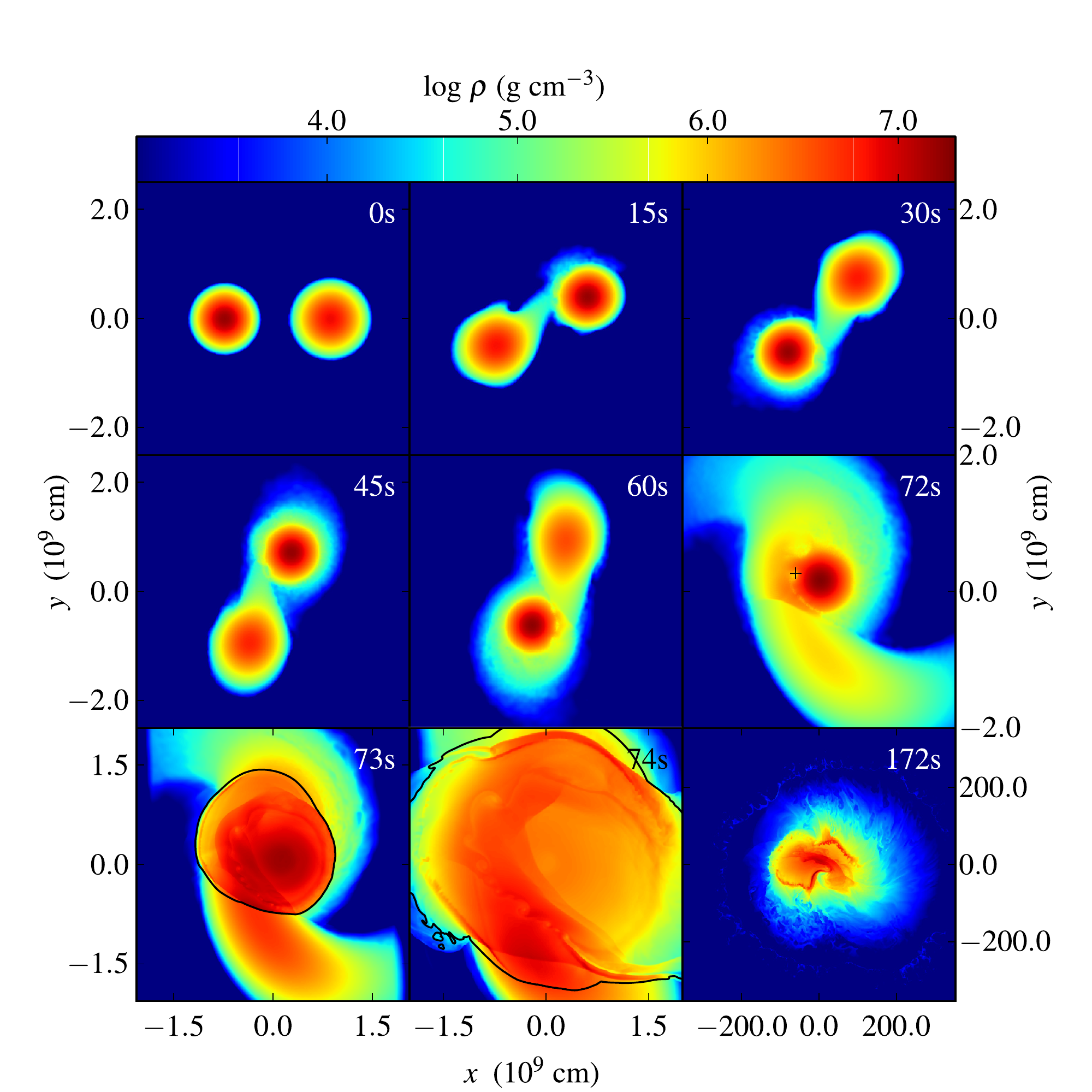}
   \caption{Time evolution of our merger model (colour-coded is the 
   logarithm of the density). Initially, the two CO WDs 
   ($0.9$\,\msun\ and $0.76$\,\msun, respectively) orbit each
   other with a period of ${\sim}36$\,s. The first six panels show
   the inspiral phase. At $t=72$\,s, the detonation is ignited
   (indicated by the crosshairs in the middle right panel).
   The bottom panels show how the detonation front (black line)
   propagates. Note that the bottom panels have different colour 
   scales, ranging from $10^{3}\,\gcc$ to $2\times10^{6}\,\gcc$,
   $10^{2}\,\gcc$ to $10^{6}\,\gcc$ and $10^{-4}\,\gcc$ to $10\,\gcc$, 
   respectively.}
   \label{fig:merger}
\end{figure*}

\cite{pakmor2011b} used the smoothed-particle hydrodynamics 
(SPH) code stellar-{\sc gadget} \citep{pakmor2012b,springel2005a}, 
to study the inspiral of various pairs of low-mass WDs with different 
mass ratios. Including a 13-isotope $\alpha$-network, they investigated 
whether the mergers of these binaries reach sufficiently high 
temperature and density that the system could detonate.
Here, we simulate the 
explosion of the lowest-mass system for which they found that a 
detonation is possible, a pair of a 0.9 and a 0.76 \msun WD\@.

We take the hydrodynamic structure of the merger
at the time of detonation initiation as obtained by \citet{pakmor2011b} 
and use the {\sc leafs} code 
\citep{reinecke2002b} to model the propagation of the detonation 
through the merged object. {\sc leafs} applies the 
level-set technique \citep{reinecke1999a} to model detonation 
fronts \citep{fink2010a}. To keep track of the rapidly streaming 
ejecta, we use the expanding-grid technique of \citet{roepke2005c}.

The full evolution of our model from the beginning of the SPH 
inspiral simulation through detonation initiation to homologous 
expansion is shown in Fig.~\ref{fig:merger}. 
At the start ($t=0$\,s), the binary system has an
orbital period of ${\sim}36$\,s on a circular co-rotating orbit. 
As shown by \citet{pakmor2012b}, this relatively tight orbit has no
significant effect on the detonation initiation since, to the accuracy 
afforded by current simulations, resolution is the limiting factor.
In the following 60\,s the primary (more massive) WD accretes
matter from the tidally deformed secondary WD\@. Owing to the high
accretion rate, material on the surface of the primary WD
is compressed and heated. In the hottest regions, C
burning is ignited, and most of the C is burned in a local 
thermonuclear runaway that reaches $T=2.67\times10^9\,\mathrm{K}$ 
and $\rho=1.91\times10^6\,\gcc$ at 72\,s (black cross 
in Fig.~\ref{fig:merger}).

We assume that a detonation forms at this instant 
\citep{seitenzahl2009b} and map the 
merger to a uniform $768^3$ Cartesian grid to model the detonation 
with {\sc leafs}. Within the mapping process $0.027$\,\msun\ are
lost since some SPH particles lie outside the selected box size of
$4\times10^9\,\mathrm{cm}$.
However, the initial density of this material is so low
that it will not be burned and neither 
affects the dynamics of the ejecta nor the synthetic observables.

On the grid the detonation is ignited in a spherical bubble of 
radius
$3\times10^7\,\mathrm{cm}$ around the hottest cell. This ignition 
regions contains 799 cells and consists mainly of O and 
intermediate-mass elements due to the preceding C burning. 
Within 2\,s after
detonation initiation (74\,s), almost all material is burned and the 
explosion ejecta are unbound (energy release from nuclear burning:
$1.4\times10^{51}$\,erg). At 172\,s (corresponding to 100\,s 
after detonation initiation) the ejecta are streaming freely with an
asymptotic kinetic energy of $1.1\times10^{51}$\,erg.

\section{Nucleosynthesis}
\label{sec:nucleosynthesis}

\begin{figure*}
   \centering
   \includegraphics[width=0.95\linewidth]{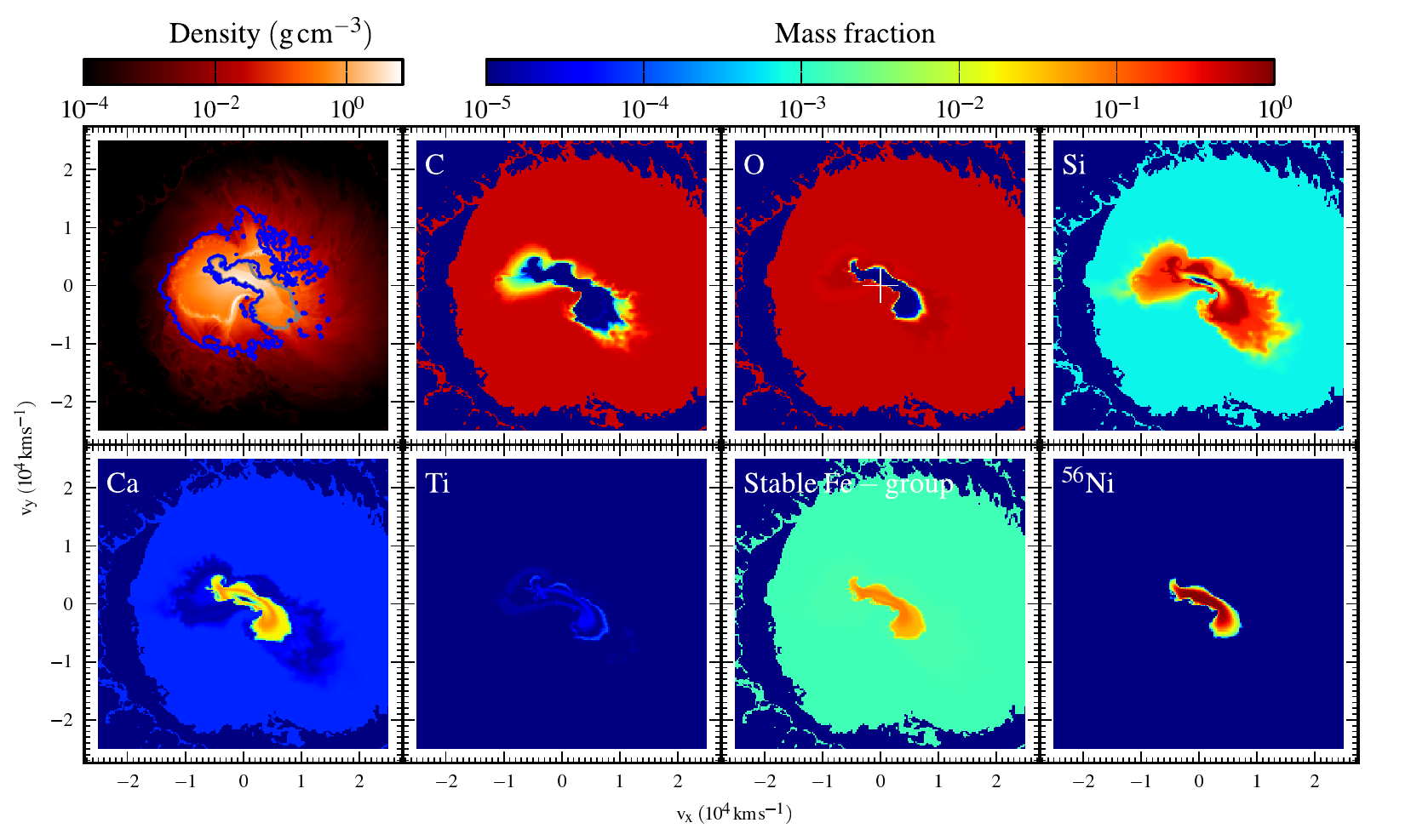}
   \caption{Slice through the midplane ($x$--$y$) of the simulation volume
     when the ejecta have reached homologous expansion at 
     $100\,\mathrm{s}$ after the explosion. We show the mass density 
     (top left) and mass fractions of select species as outlined in 
     the different panels. The contours in the density plot indicate
     regions that contain $>$90 percent material originating from
     the secondary (blue) or primary (grey) WD.}
   \label{fig:comp}
\end{figure*}

To obtain the detailed chemical composition of the explosion ejecta,
we perform a nucleosynthesis post-processing calculation with a
$384$-isotope nuclear network \citep{travaglio2004a}. As
input for this calculation we use $10^6$ Lagrangian tracer particles
for which we recorded the thermodynamic trajectories throughout the 
hydrodynamic explosion simulation. 
For the initial composition of the tracer particles we take
50\% O and 48.29\% C (by mass). Assuming that the main-sequence 
progenitor had a solar metallicity, the remaining 1.71\% are 
distributed according to the solar values of \citet{asplund2009a} 
for all elements but H and He. To account for core He-burning all 
primary C, N, O was converted to \nuc{22}{Ne} (by number).

The final ejecta contain $0.21$\,\msun\ of IGEs,
of which $0.18$\,\msun\ are \nuc{56}{Ni}. The most abundant other
species are O ($0.50$\,\msun), Si ($0.37$\,\msun), 
C ($0.21$\,\msun) and S ($0.14$\,\msun).

To obtain the spatial distribution of chemical species in the ejecta,
the final composition of the tracer particles is mapped on a $200^3$ 
Cartesian grid in asymptotic velocity space. For this, we use an
SPH-like algorithm that approximately conserves the integrated
yields (see \citealt{kromer2010a}). As can be seen from 
Fig.~\ref{fig:comp}, the distribution of the ejecta is rather 
complex, a direct consequence that the merging binary
is quite asymmetric at the time of detonation initiation (72\,s in
Fig.~\ref{fig:merger}). Since detonations propagate faster at higher
densities, the primary WD burns first, producing ashes that consist 
mainly of iron-group and intermediate-mass elements and some O\@. As 
these ashes expand, they wrap around the tidal tail, which once was 
the secondary WD\@. In this low-density material the detonation 
propagates more slowly and burning is less complete, leaving a
significant amount of unburned O\@. This O from the secondary WD 
stays at low velocities and reaches down to the centre of the ejecta.

\section{Comparison with observations}
\label{sec:observables}

\begin{figure*}
   \centering
   \includegraphics[width=0.95\linewidth]{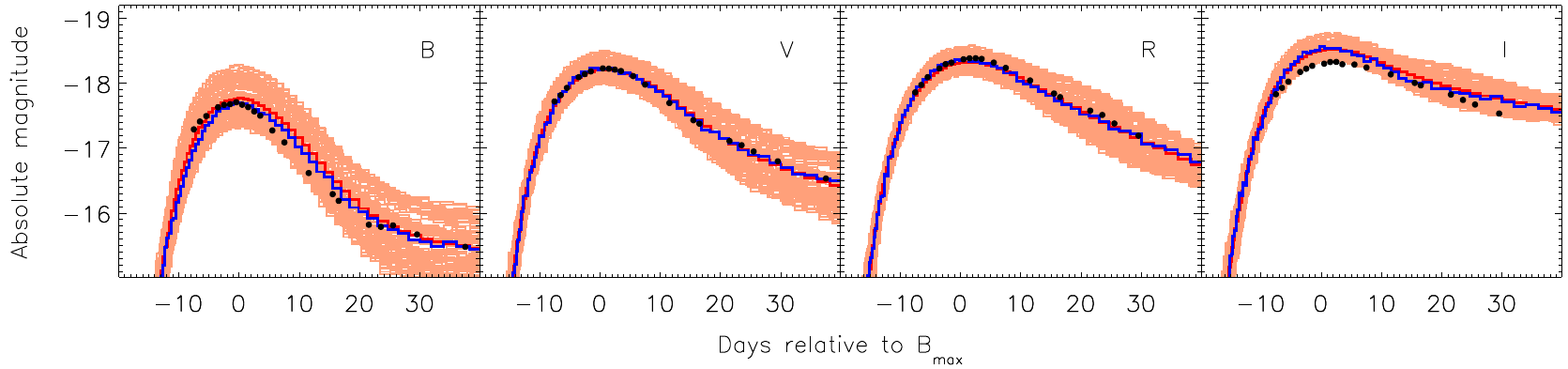}
   \caption{Broad-band lightcurves of our model for different filters
     ($B$, $V$, $R$, $I$ from left to right, respectively). While 
     100 light-red lines indicate the spread due to different viewing 
     angles, an individual line-of-sight similar to SN~2010lp is
     highlighted in blue (the dark-red line shows the angle-average). 
     Time is given relative to $B$-band maximum. 
     For comparison, the observed photometry of SN~2010lp (Pignata
     et al.\ in prep.) is overplotted as black circles.}
   \label{fig:lightcurves}
\end{figure*}

\begin{figure}
   \centering
   \includegraphics[width=\linewidth]{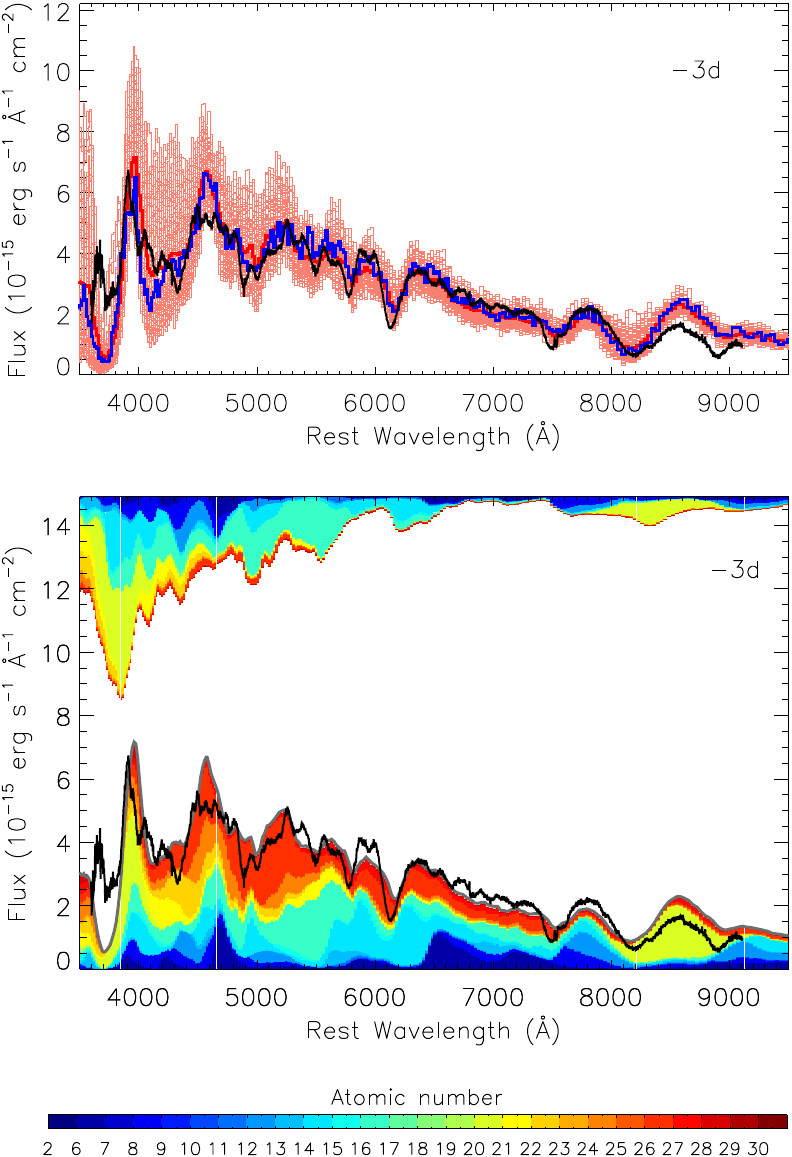}
   \caption{Top: SN~2010lp about 3 days before maximum (black) compared
      to synthetic spectra of our merger model as seen along 100
      different viewing angles (light-red lines). For comparison, an
      angle-averaged spectrum (dark red) and an individual line-of-sight
      (blue) similar to SN~2010lp are shown. 
      Bottom: angle-averaged synthetic spectrum of our merger model 
      (grey)
      and SN~2010lp (black) about 3 days before $B$-band 
      maximum. For a description of the colour coding see 
      \citep{kromer2013a}, fig. 6.}
   \label{fig:spectrum}
\end{figure}

To obtain synthetic observables for our model, we perform Monte Carlo
radiative transfer simulations with the \textsc{artis} code
\citep{kromer2009a,sim2007b}. As input for this simulation, we take 
the density distribution from the hydro model and
the composition structure from the mapping of the tracer 
particles, and rescale both to a $50^3$ Cartesian grid.
For our simulation we use the atomic data set as described
by \citet{gall2012a} and propagate $10^8$ photon packets for 111 
logarithmically spaced time steps between 2 and 120\,d after 
explosion. To reduce the
computational costs, a grey approximation is used in optically thick
cells (cf.\ \citealt{kromer2009a}), and for $t<3$\,d local
thermodynamic equilibrium is assumed.

In Fig.~\ref{fig:lightcurves} we present broad-band lightcurves of 
our model in comparison to the observed photometry of SN~2010lp 
(Pignata et al.\ in prep.). As a consequence of the asymmetric 
ejecta of our model, the lightcurves display a significant spread for
different lines-of-sight. In $B$ band, for example, we find peak 
magnitudes between $-18.25$ and $-17.30$ while \dm15 varies between 
0.95 and 1.38. In the redder bands, where the optical depths are
smaller, the viewing-angle sensitivity becomes smaller, 
since a larger fraction of the total ejecta contribute to the 
emission (compare \citealt{kromer2009a,kromer2010a}). 

Along several of the 100 different lines-of-sight, shown in 
Fig.~\ref{fig:lightcurves} (as well as angle-averaged) our 
synthetic lightcurves show excellent agreement with SN~2010lp, 
especially in $V$ and $R$. In particular, our model
naturally reproduces the low peak brightness, the 
broad lightcurves and the absence of a secondary maximum
in the $I$ band as observed in SN~2010lp. 

Fig.~\ref{fig:spectrum} shows the spectrum of our model for different
lines-of-sight three days before $B$-band maximum in comparison to 
SN~2010lp at the same epoch (Pignata et al.\ in prep.). As for the 
lightcurves, we find a significant sensitivity to viewing-angle 
effects in the spectrum. For several lines-of-sight as well as the 
angle-average our synthetic spectrum reproduces SN~2010lp remarkably
well. While there are minor discrepancies in the
strength of the \ions{Ca}{ii} (slightly too strong) and \ions{Si}{ii} 
(slightly too weak) features, our model clearly
shows the hallmark features of 1991bg-like SNe which are also found
in SN~2010lp: the broad \ions{Ti}{ii} absorption trough between 4000 
and 4400\,\AA\, the relatively strong \ions{O}{i} triplet 
$\lambda\lambda$7772,7774,7775 and low velocities in the observed
absorption features compared to normal SNe~Ia. In fact, the velocities
of our spectra are even slightly lower than those observed in SN~2010lp.

Regarding the observed [\ions{O}{i}] emission in a late-time spectrum
of SN~2010lp \citep{taubenberger2013b}, we cannot make a direct
prediction for our model at the moment. Our multi-dimensional 
radiative transfer code \textsc{artis} does not yet take into 
account the non-thermal excitation and ionization processes
necessary to model nebular spectra. One-dimensional codes do not
provide a viable alternative either, since mapping the highly 
asymmetric ejecta
of our merger model to 1D leads to artificial mixing of
chemical species, preventing any useful prediction. 
However, from Fig.~\ref{fig:comp} it is obvious
that our model has O at low velocities, a pre-requisite for 
narrow [\ions{O}{i}] emission in nebular spectra as observed 
in SN~2010lp.

\section{Discussion}
\label{sec:discussion}

\begin{figure}
   \centering
   \includegraphics[width=0.95\linewidth]{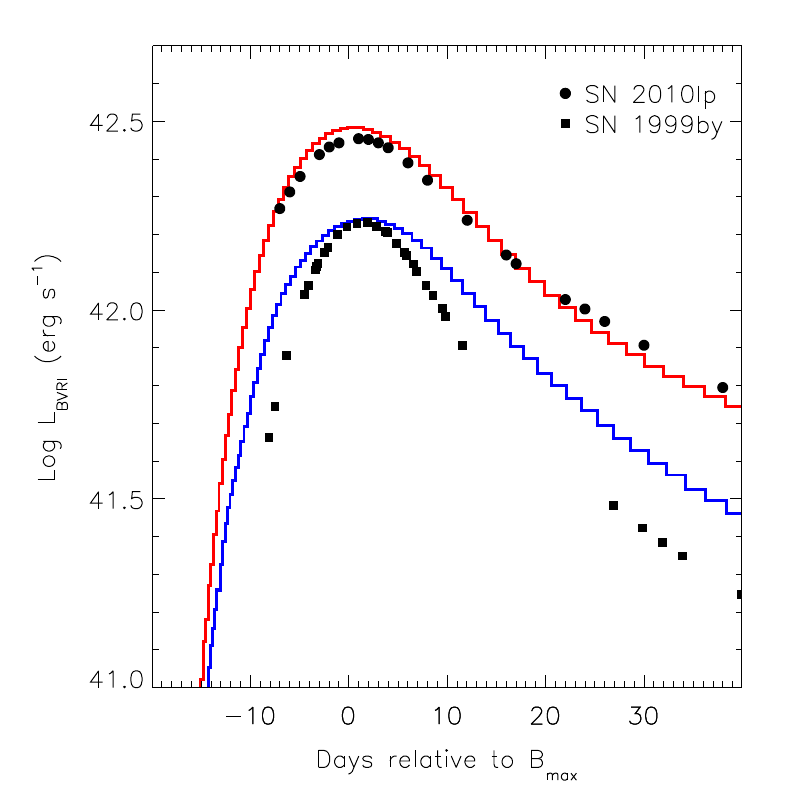}
   \caption{$BVRI$ bolometric lightcurves of our merger model (red) 
     and the 0.9+0.9 \msun\ merger (blue) of \citet{pakmor2010a}.
     For comparison, pseudobolometric lightcurves constructed from 
     $BVRI$ photometry of SNe 1999by \citep{garnavich2004a} 
     and 2010lp (Pignata et al.\ in prep.) are overplotted.}
   \label{fig:91bgcomp}
\end{figure}

In comparison to the 0.9+0.9\,\msun\ merger of \citet{pakmor2010a},
our 0.9+0.76\,\msun\ model is significantly brighter 
(Fig.~\ref{fig:91bgcomp}). This is due to an increased yield of 
\nuc{56}{Ni} in our model. As discussed by \citet{pakmor2011b}, 
the \nuc{56}{Ni} production depends only on the density profile of 
the primary WD at the time of explosion (for a given mass 
ratio $q=M_\mathrm{secondary}/M_\mathrm{primary}$ this translates to a
dependence on $M_\mathrm{primary}$). In our model, with $q=0.84$, 
the density profile of the 
primary WD at ignition is almost identical to that of an isolated, 
hydrostatic WD\@. In contrast, in the \citet{pakmor2010a} merger, 
with $q=1$, tidal interactions between primary and 
secondary are strong enough to disturb the density profiles of both 
WDs, thus reducing their central density and consequently
the \nuc{56}{Ni} production.

Owing to the low luminosity and the peculiar spectral features of
the 0.9+0.9\,\msun\ merger, \citet{pakmor2010a} proposed their model 
as a potential progenitor system for 1991bg-like SNe. However, they 
failed to reproduce the rapid lightcurve evolution typical of 
those SNe (Fig.~\ref{fig:91bgcomp}). Since the opacity in SN
ejecta is correlated with their mass, this 
indicates that the ejecta mass of the \citet{pakmor2010a} model
is too large for 1991bg-like SNe. 

\citet{maguire2011a} argued that a violent merger of 
two low-mass CO WDs, as presented by \citet{pakmor2010a}, might 
be the progenitor of PTF10ops---another 
slowly-declining subluminous SN~Ia.
The excellent agreement of our model and SN~2010lp
strengthens this connection between low-mass CO--CO WD mergers
(primary masses ${\sim}0.9$\,\msun)
and the emerging class of slowly-evolving subluminous SNe~Ia like 
PTF10ops and SN~2010lp.

While binary population synthesis calculations of \citet{ruiter2013a}
do not predict a sufficiently large number of low-mass CO--CO mergers 
to account for the 
observed number of all faint SNe~Ia (see their Fig.~5), the number 
might be sufficient 
to account for slowly-declining subluminous SNe (from the present 
observational sample their number is significantly smaller than that 
of 1991bg-like SNe).

1991bg-like systems, in contrast, require less massive ejecta.
Within the helium-ignited violent merger scenario \citep{pakmor2013a}
CO--He WD systems with a CO WD mass of ${\sim}0.8$--$0.9$\,\msun\ might 
provide appropriate progenitors for this class
of objects. According to population synthesis calculations of 
\citet{ruiter2011a}, mass-transferring CO--He systems should be 
abundant in that mass range. However, more detailed studies, in
particular explosion simulations and radiative transfer calculations,
will be required to investigate the viability of this model.

In agreement with the absence of narrow [\ions{O}{i}] lines in nebular 
spectra of SN~1991bg and SN~1999by, CO--He mergers would not be 
expected to show [\ions{O}{i}] emission at late times, since the 
central O in our model originates from the secondary WD\@. 

For CO--CO mergers the question of whether low-velocity [\ions{O}{i}] 
emission may be expected in late-time spectra is more difficult. A 
first necessary condition for narrow [\ions{O}{i}] emission is of 
course the presence of O close to the centre of the ejecta. This
depends on the mass ratio of the binary systems: mergers with 
$q\approx1$ do not produce O in the centre. In such systems
the secondary is already almost destroyed at the time the primary 
explodes. Consequently, the ejecta of the primary are able to 
push the material of the secondary away. In contrast, in mergers 
with $q<1$ the secondary is still very compact when the primary 
explodes, and thus only marginally affected by the ejecta of the 
primary. Consequently, the O-rich ashes of the secondary WD
stay in the central part of the ejecta.

However, the mere presence of O close to the centre of the ejecta 
is not sufficient for low-velocity [\ions{O}{i}] emission in nebular 
spectra. This depends on the exact excitation and ionisation state 
of the plasma (see \citealt{taubenberger2013b}
for a detailed discussion). Low-mass 
violent CO--CO mergers with central O, like our model, could show 
[\ions{O}{i}] emission due to their low luminosity (note, 
however, that the spatial proximity between \nuc{56}{Ni} and O in 
our model could still lead to ionisation of the O). In contrast,
massive mergers, like the 1.1+0.9\,\msun merger of \citet{pakmor2012a}
that resembles normal SNe~Ia but also contains O in the central ejecta, 
are more luminous. Consequently, the central O may be ionised, leading 
to no [\ions{O}{i}] emission in agreement with late-time observations
of normal SNe~Ia.

\section{Conclusions}
\label{sec:conclusions}

We have presented an explosion model of the violent merger of two 
CO WDs of 0.9 and 0.76 \msun, respectively. In the explosion ejecta
we find O close to the centre---a pre-requisite for 
narrow [\ions{O}{i}] emission in nebular spectra as observed in 
the subluminous SN~Ia 2010lp. Moreover, from radiative transfer 
simulations we have shown that our model adequately reproduces the
observed early-time observables of SN~2010lp---in particular its 
low luminosity and slowly-evolving lightcurves, but also its
colours and peculiar spectral features.

Our findings support the suggestion of 
\citet{maguire2011a} that violent mergers of two CO WDs 
with low-mass primaries (${\sim}0.9$\,\msun) are a better
match to the emerging class of slowly-evolving subluminous 
SNe~Ia like PTF10ops and SN~2010lp, rather than 
to the rapidly declining 1991bg-like SNe as suggested by 
\citet{pakmor2010a}.

The excellent agreement between our explosion model and SN~2010lp
demonstrates the power of theoretical explosion models in finding 
possible progenitors of (peculiar) SNe~Ia. In particular the merger 
scenario, comprising a large variety of possible progenitor properties
warrants detailed scrutiny.

\begin{acknowledgements}This work was supported by the Deutsche
  Forschungs\-gemeinschaft via TRR-33 ``The Dark Universe'', the
  graduate school GRK-1147 ``Theoretical Astrophysics and Particle
  Physics'' and the Emmy Noether Program (RO 3676/1-1). F.K.R.\ is
  supported by the ARCHES prize of the German Ministry of Education \&
  Research, R.P.\ by the European Research Council (ERC-StG
  EXAGAL-308037) and G.P.\ from Millennium Center for Supernova
  Science (P10-064-F), with input from Fondo de Innovaci\'on para la
  Competitividad, del Ministerio de Economia, Fomento y Turismo de
  Chile.  The simulations were carried out at the J\"{u}lich
  Supercomputing Centre (project hmu14).\end{acknowledgements}


\begin{thebibliography}{46}
\expandafter\ifx\csname natexlab\endcsname\relax\def\natexlab#1{#1}\fi

\bibitem[{{Asplund} {et~al.}(2009){Asplund}, {Grevesse}, {Sauval}, \&
  {Scott}}]{asplund2009a}
{Asplund}, M., {Grevesse}, N., {Sauval}, A.~J., \& {Scott}, P. 2009, \araa, 47,
  481

\bibitem[{{Bloom} {et~al.}(2012){Bloom}, {Kasen}, {Shen}, {Nugent}, {Butler},
  {Graham}, {Howell}, {Kolb}, {Holmes}, {Haswell}, {Burwitz}, {Rodriguez}, \&
  {Sullivan}}]{bloom2012a}
{Bloom}, J.~S., {Kasen}, D., {Shen}, K.~J., {et~al.} 2012, \apjl, 744, L17

\bibitem[{{Dilday} {et~al.}(2012){Dilday}, {Howell}, {Cenko}, {Silverman},
  {Nugent}, {Sullivan}, {Ben-Ami}, {Bildsten}, {Bolte}, {Endl}, {Filippenko},
  {Gnat}, {Horesh}, {Hsiao}, {Kasliwal}, {Kirkman}, {Maguire}, {Marcy},
  {Moore}, {Pan}, {Parrent}, {Podsiadlowski}, {Quimby}, {Sternberg}, {Suzuki},
  {Tytler}, {Xu}, {Bloom}, {Gal-Yam}, {Hook}, {Kulkarni}, {Law}, {Ofek},
  {Polishook}, \& {Poznanski}}]{dilday2012a}
{Dilday}, B., {Howell}, D.~A., {Cenko}, S.~B., {et~al.} 2012, Science, 337, 942

\bibitem[{{Filippenko} {et~al.}(1992){Filippenko}, {Richmond}, {Branch},
  {Gaskell}, {Herbst}, {Ford}, {Treffers}, {Matheson}, {Ho}, {Dey}, {Sargent},
  {Small}, \& {van Breugel}}]{filippenko1992b}
{Filippenko}, A.~V., {Richmond}, M.~W., {Branch}, D., {et~al.} 1992, \aj, 104,
  1543

\bibitem[{{Fink} {et~al.}(2013){Fink}, {Kromer}, {Seitenzahl},
  {Ciaraldi-Schoolmann}, {Roepke}, {Sim}, {Pakmor}, {Ruiter}, \&
  {Hillebrandt}}]{fink2013a}
{Fink}, M., {Kromer}, M., {Seitenzahl}, I.~R., {et~al.} 2013, ArXiv e-prints

\bibitem[{{Fink} {et~al.}(2010){Fink}, {R{\"o}pke}, {Hillebrandt},
  {Seitenzahl}, {Sim}, \& {Kromer}}]{fink2010a}
{Fink}, M., {R{\"o}pke}, F.~K., {Hillebrandt}, W., {et~al.} 2010, \aap, 514,
  A53

\bibitem[{{Gall} {et~al.}(2012){Gall}, {Taubenberger}, {Kromer}, {Sim},
  {Benetti}, {Blanc}, {Elias-Rosa}, \& {Hillebrandt}}]{gall2012a}
{Gall}, E.~E.~E., {Taubenberger}, S., {Kromer}, M., {et~al.} 2012, \mnras, 427,
  994

\bibitem[{{Garnavich} {et~al.}(2004){Garnavich}, {Bonanos}, {Krisciunas},
  {Jha}, {Kirshner}, {Schlegel}, {Challis}, {Macri}, {Hatano}, {Branch},
  {Bothun}, \& {Freedman}}]{garnavich2004a}
{Garnavich}, P.~M., {Bonanos}, A.~Z., {Krisciunas}, K., {et~al.} 2004, \apj,
  613, 1120

\bibitem[{{Gonz{\'a}lez Hern{\'a}ndez} {et~al.}(2012){Gonz{\'a}lez
  Hern{\'a}ndez}, {Ruiz-Lapuente}, {Tabernero}, {Montes}, {Canal},
  {M{\'e}ndez}, \& {Bedin}}]{hernandez2012a}
{Gonz{\'a}lez Hern{\'a}ndez}, J.~I., {Ruiz-Lapuente}, P., {Tabernero}, H.~M.,
  {et~al.} 2012, \nat, 489, 533

\bibitem[{{Hillebrandt} {et~al.}(2013){Hillebrandt}, {Kromer}, {R{\"o}pke}, \&
  {Ruiter}}]{hillebrandt2013a}
{Hillebrandt}, W., {Kromer}, M., {R{\"o}pke}, F.~K., \& {Ruiter}, A.~J. 2013,
  Frontiers of Physics, 8, 116

\bibitem[{{Hillebrandt} \& {Niemeyer}(2000)}]{hillebrandt2000a}
{Hillebrandt}, W. \& {Niemeyer}, J.~C. 2000, \araa, 38, 191

\bibitem[{{H{\"o}flich} \& {Khokhlov}(1996)}]{hoeflich1996a}
{H{\"o}flich}, P. \& {Khokhlov}, A. 1996, \apj, 457, 500

\bibitem[{{Jordan} {et~al.}(2012){Jordan}, {Perets}, {Fisher}, \& {van
  Rossum}}]{jordan2012b}
{Jordan}, IV, G.~C., {Perets}, H.~B., {Fisher}, R.~T., \& {van Rossum}, D.~R.
  2012, \apjl, 761, L23

\bibitem[{{Kasen} {et~al.}(2009){Kasen}, {R{\"o}pke}, \&
  {Woosley}}]{kasen2009a}
{Kasen}, D., {R{\"o}pke}, F.~K., \& {Woosley}, S.~E. 2009, \nat, 460, 869

\bibitem[{{Kozma} {et~al.}(2005){Kozma}, {Fransson}, {Hillebrandt},
  {Travaglio}, {Sollerman}, {Reinecke}, {R{\"o}pke}, \&
  {Spyromilio}}]{kozma2005a}
{Kozma}, C., {Fransson}, C., {Hillebrandt}, W., {et~al.} 2005, \aap, 437, 983

\bibitem[{{Kromer} {et~al.}(2013){Kromer}, {Fink}, {Stanishev}, {Taubenberger},
  {Ciaraldi-Schoolman}, {Pakmor}, {R{\"o}pke}, {Ruiter}, {Seitenzahl}, {Sim},
  {Blanc}, {Elias-Rosa}, \& {Hillebrandt}}]{kromer2013a}
{Kromer}, M., {Fink}, M., {Stanishev}, V., {et~al.} 2013, \mnras, 429, 2287

\bibitem[{{Kromer} \& {Sim}(2009)}]{kromer2009a}
{Kromer}, M. \& {Sim}, S.~A. 2009, \mnras, 398, 1809

\bibitem[{{Kromer} {et~al.}(2010){Kromer}, {Sim}, {Fink}, {R{\"o}pke},
  {Seitenzahl}, \& {Hillebrandt}}]{kromer2010a}
{Kromer}, M., {Sim}, S.~A., {Fink}, M., {et~al.} 2010, \apj, 719, 1067

\bibitem[{{Leibundgut} {et~al.}(1993){Leibundgut}, {Kirshner}, {Phillips},
  {Wells}, {Suntzeff}, {Hamuy}, {Schommer}, {Walker}, {Gonzalez}, {Ugarte},
  {Williams}, {Williger}, {Gomez}, {Marzke}, {Schmidt}, {Whitney}, {Coldwell},
  {Peters}, {Chaffee}, {Foltz}, {Rehner}, {Siciliano}, {Barnes}, {Cheng},
  {Hintzen}, {Kim}, {Maza}, {Parker}, {Porter}, {Schmidtke}, \&
  {Sonneborn}}]{leibundgut1993a}
{Leibundgut}, B., {Kirshner}, R.~P., {Phillips}, M.~M., {et~al.} 1993, \aj,
  105, 301

\bibitem[{{Li} {et~al.}(2011){Li}, {Bloom}, {Podsiadlowski}, {Miller}, {Cenko},
  {Jha}, {Sullivan}, {Howell}, {Nugent}, {Butler}, {Ofek}, {Kasliwal},
  {Richards}, {Stockton}, {Shih}, {Bildsten}, {Shara}, {Bibby}, {Filippenko},
  {Ganeshalingam}, {Silverman}, {Kulkarni}, {Law}, {Poznanski}, {Quimby},
  {McCully}, {Patel}, {Maguire}, \& {Shen}}]{li2011b}
{Li}, W., {Bloom}, J.~S., {Podsiadlowski}, P., {et~al.} 2011, \nat, 480, 348

\bibitem[{{Livne}(1990)}]{livne1990a}
{Livne}, E. 1990, \apjl, 354, L53

\bibitem[{{Ma} {et~al.}(2013){Ma}, {Woosley}, {Malone}, {Almgren}, \&
  {Bell}}]{ma2013a}
{Ma}, H., {Woosley}, S.~E., {Malone}, C.~M., {Almgren}, A., \& {Bell}, J. 2013,
  \apj, 771, 58

\bibitem[{{Maguire} {et~al.}(2011){Maguire}, {Sullivan}, {Thomas}, {Nugent},
  {Howell}, {Gal-Yam}, {Arcavi}, {Ben-Ami}, {Blake}, {Botyanszki}, {Buton},
  {Cooke}, {Ellis}, {Hook}, {Kasliwal}, {Pan}, {Pereira}, {Podsiadlowski},
  {Sternberg}, {Suzuki}, {Xu}, {Yaron}, {Bloom}, {Cenko}, {Kulkarni}, {Law},
  {Ofek}, {Poznanski}, \& {Quimby}}]{maguire2011a}
{Maguire}, K., {Sullivan}, M., {Thomas}, R.~C., {et~al.} 2011, \mnras, 418, 747

\bibitem[{{Nomoto}(1982)}]{nomoto1982b}
{Nomoto}, K. 1982, \apj, 257, 780

\bibitem[{{Pakmor} {et~al.}(2012{\natexlab{a}}){Pakmor}, {Edelmann},
  {R{\"o}pke}, \& {Hillebrandt}}]{pakmor2012b}
{Pakmor}, R., {Edelmann}, P., {R{\"o}pke}, F.~K., \& {Hillebrandt}, W.
  2012{\natexlab{a}}, \mnras, 424, 2222

\bibitem[{{Pakmor} {et~al.}(2011){Pakmor}, {Hachinger}, {R{\"o}pke}, \&
  {Hillebrandt}}]{pakmor2011b}
{Pakmor}, R., {Hachinger}, S., {R{\"o}pke}, F.~K., \& {Hillebrandt}, W. 2011,
  \aap, 528, A117+

\bibitem[{{Pakmor} {et~al.}(2010){Pakmor}, {Kromer}, {R{\"o}pke}, {Sim},
  {Ruiter}, \& {Hillebrandt}}]{pakmor2010a}
{Pakmor}, R., {Kromer}, M., {R{\"o}pke}, F.~K., {et~al.} 2010, \nat, 463, 61

\bibitem[{{Pakmor} {et~al.}(2012{\natexlab{b}}){Pakmor}, {Kromer},
  {Taubenberger}, {Sim}, {R{\"o}pke}, \& {Hillebrandt}}]{pakmor2012a}
{Pakmor}, R., {Kromer}, M., {Taubenberger}, S., {et~al.} 2012{\natexlab{b}},
  \apjl, 747, L10

\bibitem[{{Pakmor} {et~al.}(2013){Pakmor}, {Kromer}, {Taubenberger}, \&
  {Springel}}]{pakmor2013a}
{Pakmor}, R., {Kromer}, M., {Taubenberger}, S., \& {Springel}, V. 2013, \apjl,
  770, L8

\bibitem[{{Patat} {et~al.}(2007){Patat}, {Chandra}, {Chevalier}, {Justham},
  {Podsiadlowski}, {Wolf}, {Gal-Yam}, {Pasquini}, {Crawford}, {Mazzali},
  {Pauldrach}, {Nomoto}, {Benetti}, {Cappellaro}, {Elias-Rosa}, {Hillebrandt},
  {Leonard}, {Pastorello}, {Renzini}, {Sabbadin}, {Simon}, \&
  {Turatto}}]{patat2007a}
{Patat}, F., {Chandra}, P., {Chevalier}, R., {et~al.} 2007, Science, 317, 924

\bibitem[{{Reinecke} {et~al.}(2002){Reinecke}, {Hillebrandt}, \&
  {Niemeyer}}]{reinecke2002b}
{Reinecke}, M., {Hillebrandt}, W., \& {Niemeyer}, J.~C. 2002, \aap, 386, 936

\bibitem[{{Reinecke} {et~al.}(1999){Reinecke}, {Hillebrandt}, {Niemeyer},
  {Klein}, \& {Gr{\"o}bl}}]{reinecke1999a}
{Reinecke}, M., {Hillebrandt}, W., {Niemeyer}, J.~C., {Klein}, R., \&
  {Gr{\"o}bl}, A. 1999, \aap, 347, 724

\bibitem[{{R{\"o}pke}(2005)}]{roepke2005c}
{R{\"o}pke}, F.~K. 2005, \aap, 432, 969

\bibitem[{{R{\"o}pke} {et~al.}(2012){R{\"o}pke}, {Kromer}, {Seitenzahl},
  {Pakmor}, {Sim}, {Taubenberger}, {Ciaraldi-Schoolmann}, {Hillebrandt},
  {Aldering}, {Antilogus}, {Baltay}, {Benitez-Herrera}, {Bongard}, {Buton},
  {Canto}, {Cellier-Holzem}, {Childress}, {Chotard}, {Copin}, {Fakhouri},
  {Fink}, {Fouchez}, {Gangler}, {Guy}, {Hachinger}, {Hsiao}, {Chen},
  {Kerschhaggl}, {Kowalski}, {Nugent}, {Paech}, {Pain}, {Pecontal}, {Pereira},
  {Perlmutter}, {Rabinowitz}, {Rigault}, {Runge}, {Saunders}, {Smadja},
  {Suzuki}, {Tao}, {Thomas}, {Tilquin}, \& {Wu}}]{roepke2012a}
{R{\"o}pke}, F.~K., {Kromer}, M., {Seitenzahl}, I.~R., {et~al.} 2012, \apjl,
  750, L19

\bibitem[{{Ruiter} {et~al.}(2011){Ruiter}, {Belczynski}, {Sim}, {Hillebrandt},
  {Fryer}, {Fink}, \& {Kromer}}]{ruiter2011a}
{Ruiter}, A.~J., {Belczynski}, K., {Sim}, S.~A., {et~al.} 2011, \mnras, 417,
  408

\bibitem[{{Ruiter} {et~al.}(2013){Ruiter}, {Sim}, {Pakmor}, {Kromer},
  {Seitenzahl}, {Belczynski}, {Fink}, {Herzog}, {Hillebrandt}, {R{\"o}pke}, \&
  {Taubenberger}}]{ruiter2013a}
{Ruiter}, A.~J., {Sim}, S.~A., {Pakmor}, R., {et~al.} 2013, \mnras, 429, 1425

\bibitem[{{Schaefer} \& {Pagnotta}(2012)}]{schaefer2012a}
{Schaefer}, B.~E. \& {Pagnotta}, A. 2012, \nat, 481, 164

\bibitem[{{Seitenzahl} {et~al.}(2013){Seitenzahl}, {Ciaraldi-Schoolmann},
  {R{\"o}pke}, {Fink}, {Hillebrandt}, {Kromer}, {Pakmor}, {Ruiter}, {Sim}, \&
  {Taubenberger}}]{seitenzahl2013a}
{Seitenzahl}, I.~R., {Ciaraldi-Schoolmann}, F., {R{\"o}pke}, F.~K., {et~al.}
  2013, \mnras, 429, 1156

\bibitem[{{Seitenzahl} {et~al.}(2009){Seitenzahl}, {Meakin}, {Townsley},
  {Lamb}, \& {Truran}}]{seitenzahl2009b}
{Seitenzahl}, I.~R., {Meakin}, C.~A., {Townsley}, D.~M., {Lamb}, D.~Q., \&
  {Truran}, J.~W. 2009, \apj, 696, 515

\bibitem[{{Shappee} {et~al.}(2013){Shappee}, {Stanek}, {Pogge}, \&
  {Garnavich}}]{shappee2013a}
{Shappee}, B.~J., {Stanek}, K.~Z., {Pogge}, R.~W., \& {Garnavich}, P.~M. 2013,
  \apjl, 762, L5

\bibitem[{{Sim}(2007)}]{sim2007b}
{Sim}, S.~A. 2007, \mnras, 375, 154

\bibitem[{{Sim} {et~al.}(2010){Sim}, {R{\"o}pke}, {Hillebrandt}, {Kromer},
  {Pakmor}, {Fink}, {Ruiter}, \& {Seitenzahl}}]{sim2010a}
{Sim}, S.~A., {R{\"o}pke}, F.~K., {Hillebrandt}, W., {et~al.} 2010, \apjl, 714,
  L52

\bibitem[{{Springel}(2005)}]{springel2005a}
{Springel}, V. 2005, \mnras, 364, 1105

\bibitem[{{Taubenberger} {et~al.}(2008){Taubenberger}, {Hachinger}, {Pignata},
  {Mazzali}, {Contreras}, {Valenti}, {Pastorello}, {Elias-Rosa},
  {B{\"a}rnbantner}, {Barwig}, {Benetti}, {Dolci}, {Fliri}, {Folatelli},
  {Freedman}, {Gonzalez}, {Hamuy}, {Krzeminski}, {Morrell}, {Navasardyan},
  {Persson}, {Phillips}, {Ries}, {Roth}, {Suntzeff}, {Turatto}, \&
  {Hillebrandt}}]{taubenberger2008a}
{Taubenberger}, S., {Hachinger}, S., {Pignata}, G., {et~al.} 2008, \mnras, 385,
  75

\bibitem[{{Taubenberger} {et~al.}(2013){Taubenberger}, {Kromer}, {Pakmor},
  {Pignata}, {Maeda}, {Hachinger}, {Leibundgut}, \&
  {Hillebrandt}}]{taubenberger2013b}
{Taubenberger}, S., {Kromer}, M., {Pakmor}, R., {et~al.} 2013, \apjl, 775, L43

\bibitem[{{Travaglio} {et~al.}(2004){Travaglio}, {Hillebrandt}, {Reinecke}, \&
  {Thielemann}}]{travaglio2004a}
{Travaglio}, C., {Hillebrandt}, W., {Reinecke}, M., \& {Thielemann}, F.-K.
  2004, \aap, 425, 1029

\end{thebibliography}

\end{document}